\documentclass[aps, preprint, 12pt]{revtex4}

\usepackage{epsfig, amssymb}
\usepackage[normalem]{ulem}

\usepackage{graphicx}
\usepackage{subfigure}
\usepackage{natbib}
\usepackage{color}
\usepackage{epstopdf}

\def\bchange#1{#1}

\begin{document}

\title{On the analogy between streamlined magnetic and solid obstacles}

    \author{E.V. Votyakov}
    \email{votyakov@ucy.ac.cy, karaul@gmail.com}
    \author{S.C. Kassinos}
    \email{kassinos@ucy.ac.cy}

    \affiliation{Computational Science Laboratory UCY-CompSci, Department of
    Mechanical and Manufacturing Engineering, University of Cyprus, 75
    Kallipoleos, Nicosia 1678, Cyprus}

\begin{abstract}

Analogies are elaborated in the qualitative description of two
systems: the magnetohydrodynamic (MHD) flow moving through a
region where an external local magnetic field (magnetic obstacle)
is applied, and the ordinary hydrodynamic flow around a solid
obstacle. The former problem is of interest both practically and
theoretically, and the latter one is a classical problem being
well understood in ordinary hydrodynamics.  The first analogy is
the formation in the MHD flow of an impenetrable region -- core of
the magnetic obstacle -- as the interaction parameter $N$, i.e.
strength of the applied magnetic field, increases significantly.
The core of the magnetic obstacle is streamlined both by the
upstream flow and by the induced cross stream electric currents,
like a foreign insulated insertion placed inside the ordinary
hydrodynamic flow. In the core, closed streamlines of the mass
flow resemble contour lines of electric potential, while closed
streamlines of the electric current resemble contour lines of
pressure. The second analogy is the breaking away of attached
vortices from the recirculation pattern produced by the magnetic
obstacle when the Reynolds number $Re$, i.e. velocity of the
upstream flow, is larger than a critical value. This breaking away
of vortices from the magnetic obstacle is similar to that
occurring past a solid obstacle. Depending on the inlet and/or
initial conditions, the observed vortex shedding can be either
symmetric or asymmetric.

\end{abstract}


\maketitle

\section*{Introduction} \label{sec:intro}

External magnetic fields are heavily exploited in many practical
applications \cite{Davidson:Review:1999}, such as electromagnetic
stirring, electromagnetic brakes, and non-contact flow
measurements \cite{Thess:Votyakov:Kolesnikov:2006}. The crucial
aspect in the above applications is the Lorentz force produced by
the interaction of an external magnetic field with induced
electric currents. The currents appear because an electrically
conducting fluid moves relative to the external field. The Lorentz
force has a double effect on the flow: it suppresses turbulent
fluctuations when the intensity of the external field is strong
and spatially uniform, \bchange{but also is able to produce
vorticity} if the intensity varies spatially. If the external
magnetic field is localized in space, i.e. it acts on a finite region
of flow, then the flow is decelerated in this region and one can
say that the local magnetic field produces a virtual obstacle,
called a magnetic obstacle. Both a solid and magnetic obstacle
have a real physical effect in the sense that they impede the
flow.

The retarding effect of the external \bchange{nonuniform} magnetic
field on the liquid metal flow is well-known and has been
intensively studied in the past, see for instance books
\cite{Tananaev:book:1979, Moreau:book:1990, Davidson:book:2001,
Mueller:Buehler:Book:2001}. The overwhelming majority of works
were performed on liquid metal flows in ducts subject to fringing
magnetic fields. The main goal was to study the so-called M-shaped
velocity profile formed by directing the flow into the region of
the fringing magnetic field. The M-shaped profile is characterized
by two side jets around a central stagnant region.

\begin{figure}
\begin{center}
    \includegraphics[width=15cm, angle=0, clip=yes]{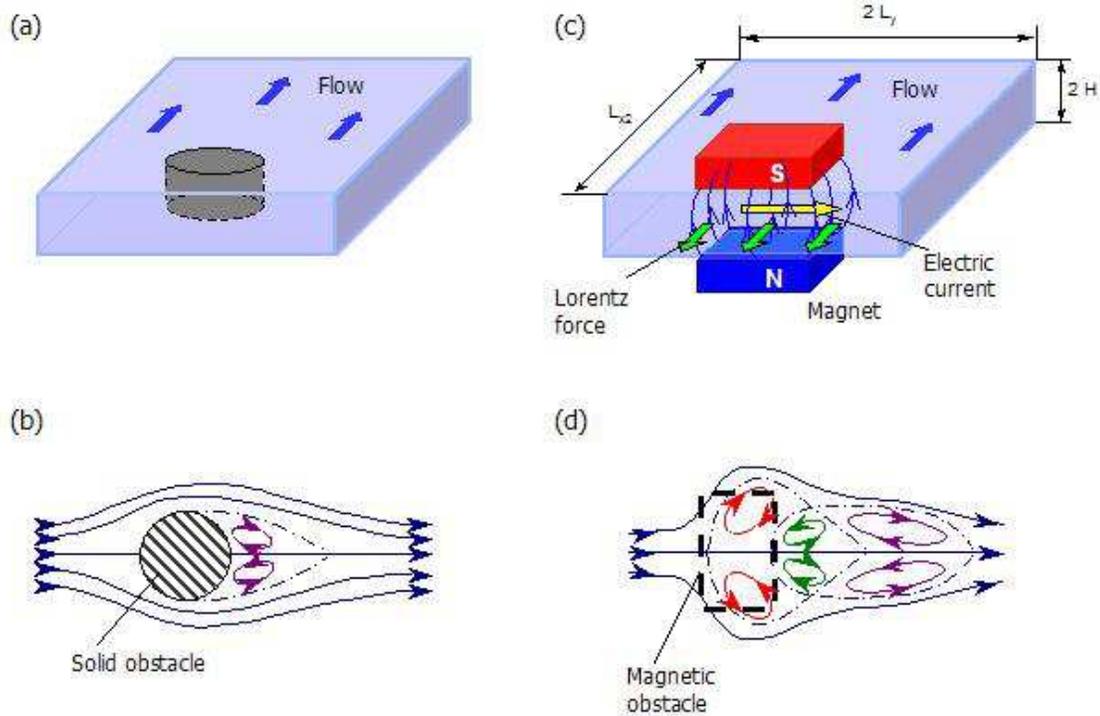}
\end{center}
    \caption{\label{Fig:Introductory}Flow around a solid
    ($a,b$) and magnetic ($c,d$) obstacle. Wake of the solid obstacle ($b$)
    shows two attached vortices, wake of the magnetic obstacle ($d$)
    shows inner magnetic (first),connecting (second) and attached
    vortices (third pair).}
\end{figure}

The flow around a solid obstacle, such as a circular cylinder
schematically given in Fig.~\ref{Fig:Introductory}($a$), is a
classical hydrodynamical problem that is qualitatively well
understood. The structure of the wake of the cylinder depends on
the Reynolds parameter $Re=u_0d/\nu$, where $u_0$ is velocity at
infinity, $d$ is the cylinder diameter, and $\nu$ the kinematic
viscosity of the fluid. Physically, $Re$ expresses the ratio of
inertial to viscous forces. When the inertia of the flow
increases, two attached vortices appear past the cylinder,
Fig.~\ref{Fig:Introductory}($b$). As the inertia of the flow
increases further, the vortices detach from the cylinder and form
the von Karman vortex street.

The flow around a magnetic obstacle like those schematically given
in Fig.~\ref{Fig:Introductory}($c$) is a rather new MHD problem
that is not yet completely understood. The first studies devoted
to liquid metal flow around a magnetic obstacle were carried out
in the former Soviet Union\cite{Gelfgat:Peterson:Sherbinin:1978,
Gelfgat:Olshanskii:1978}. 2D numerical calculations
\cite{Gelfgat:Peterson:Sherbinin:1978} have found two vortices
inside the magnetic obstacle, however, especially designed
experiments \cite{Gelfgat:Olshanskii:1978} did not confirm the
numerical finding. Lately the term `magnetic obstacle' has been
revived for Western readers in 2D numerical works
\cite{Cuevas:Smolentsev:Abdou:Pamir:2005,
Cuevas:Smolentsev:Abdou:PRE:2006, Cuevas:Smolentsev:Abdou:2006},
where authors also have found a vortex dipole in a creeping MHD
flow \cite{Cuevas:Smolentsev:Abdou:PRE:2006} and claimed that
vortex generation past a magnetic obstacle is similar to that past
a solid obstacle \cite{Cuevas:Smolentsev:Abdou:2006}.

The most recent results for the flow around a magnetic obstacle
were obtained by means of 3D numerics and physical experiments
\cite{Votyakov:PRL:2007, Votyakov:JFM:2007}. It  turns out that
the structure of the wake of the magnetic obstacle is more complex
that that of the solid obstacle. In addition to the Reynolds
number, $Re=u_0\,H/\nu$, an MHD flow is characterized by the
magnetic interaction parameter, $N=\sigma H\,B_0^2/\rho u_0$,
where $H$, $u_0$, $B_0$ are the characteristic length scale,
velocity and intensity of the applied magnetic field, and $\rho$
and $\sigma$ are the density and electric conductivity of the
fluid, see e.g.
\cite{Shercliff:book:1962,Roberts:1967,Moreau:book:1990,
Davidson:book:2001}. $N$ represents the ratio of the Lorentz force
to the inertial force. Depending on $Re$ and $N$, i.e. on the
relationship between viscous, Lorentz and inertial forces, the
liquid metal flow shows three different regimes: (1) no vortices,
when the viscous force prevails at the small Lorentz force limit,
(2) one pair of magnetic vortices when Lorentz force is high and
inertia is small, and (3) three pairs (namely, magnetic,
connecting, and attached vortices) when the Lorentz and inertial
forces dominate the viscous force. The latter case is shown in
Fig.~\ref{Fig:Introductory}$d$. We believe that this scenario for
the wake of the magnetic obstacle is the generic one and devote
the last Section of the paper to explain in detail why this is so.

\bchange{An analogy between solid and  magnetic obstacles had been
suggested from the beginning of MHD liquid metal works in the
former USSR \footnote{\bchange{Yuri Kolesnikov, coauthor of
\cite{Votyakov:PRL:2007, Votyakov:JFM:2007} used ''magnetic
obstacle'' as a working term in the 1970s in Riga, MHD center of
the former USSR, in order to stress the analogy with a solid
obstacle.}}}. Based on 2D inertialess simulations, Gelfgat
\textit{et al.} \cite{Gelfgat:Peterson:Sherbinin:1978} remarked
that a vortex dipole inside the magnetic obstacle is similar to
attached vortices past a solid obstacle.  Afterwards, a series of
experiments of Gelfgat \textit{et al.}
\cite{Gelfgat:Olshanskii:1978} failed to confirm the numerical
results and this let them to question the original suggestion. As
has been shown recently \cite{Votyakov:PRL:2007}, however, the
problem with the suggested analogy is that the numerically
observed flow structures were magnetic vortices fixed inside the
magnetic obstacle rather than the attached vortices disposed past
the magnetic obstacle. The 2D numerical results
\cite{Gelfgat:Peterson:Sherbinin:1978} were correct for creeping
flow, while the physical experiments \cite{Gelfgat:Olshanskii:1978}
were performed at high Reynolds numbers $Re$, and so they
failed to produce any vortices since the interaction parameter $N$
was not high enough \cite{Votyakov:JFM:2007}.

\bchange{Then, Cuevas \textit{et al.}
\cite{Cuevas:Smolentsev:Abdou:Pamir:2005,
Cuevas:Smolentsev:Abdou:2006}, by means of 2D and quasi-2D
numerics \footnote{\bchange{Fig. 4 in
\cite{Cuevas:Smolentsev:Abdou:Pamir:2005} and Fig.4 in
\cite{Cuevas:Smolentsev:Abdou:2006} are very similar despite the
fact that the paper \cite{Cuevas:Smolentsev:Abdou:Pamir:2005} is a
 2D simulation while the paper
\cite{Cuevas:Smolentsev:Abdou:2006} is a quasi-2D approach with a
friction term. One may conclude that the friction term is of minor
significance in 2D models.}}}, concluded that the vortex
generation past a magnetic obstacle is similar to that past a
solid cylinder. Although this conclusion is correct in general, it
is rather obvious: any decelerating force generates vorticity that
is then translated downstream, exciting in the process the von
Karman vortex street. Moreover, at high Reynolds numbers, 2D
numerics is not a suitable method to analyze the flow around the
magnetic obstacle because it neglects the Hartmann friction
\cite{Votyakov:JFM:2007}, and as a result, fails to describe the
stable six-vortex structure shown in
Fig.~\ref{Fig:Introductory}$d$ and discussed later.

The possible reason why the previous two analogies were either
imprecise \cite{Gelfgat:Peterson:Sherbinin:1978} or trivially
correct \cite{Cuevas:Smolentsev:Abdou:2006} is because the
previous interpretations were not based on full 3D numerical
simulations at high Reynolds numbers. As a result, the previous
interpretations suffered from the lack of a concrete and clear
demonstration. The basic message of our paper is to report the
correct, in our opinion, analogy, and to confirm it by means of
concrete 3D numerical results.

A fruitful way of thinking about the  similarity between magnetic
and solid obstacles is that the magnetic and connecting vortices,
taken together as one entity, form the body of a virtual insertion
in the MHD flow \cite{Votyakov:PRL:2007}. One can understand it by
recalling the classical potential flow theory. In this theory, a
real streamlined cylinder is modelled \bchange{by a virtual
imaginable} vortex dipole. In the MHD case, we have an opposite
picture: a magnetic obstacle, \bchange{that can be understood as a
virtual bluff body}, manifests itself by means of real physical
vortices. The present paper supports this idea with two new
aspects.

The first is the impenetrable core of the magnetic obstacle. It
originates in the center of the magnetic gap as the magnetic
interaction parameter $N$ increases. When $N$ is very large, both
mass transfer and electric field vanish in the region between
magnetic poles. This region looks as if frozen by the external
magnetic field so that the upstream flow and crosswise electric
currents can not penetrate inside it. Thus, the core of the
magnetic obstacle is similar to an insulated solid obstacle inside
an ordinary hydrodynamical flow with crosswise electric currents
and \textit{without} an external magnetic field. In this latter
case, because of the absence of a magnetic field, the crosswise
electric currents go around the insulated insertion without
affecting the mass flow. Magnetic vortices are located aside the
core and compensate shear stresses, like a ball-bearing between the
impenetrable region and upstream flow.

At first glance, the appearance of the core of the magnetic
obstacle can be admitted as intensively studied before. Indeed, a
stagnant region between two side jets is well-known for duct flows
subject to fringing magnetic fields. However, at a closer
examination one finds that the fringing magnetic field is not the
case of the magnetic obstacle. In the former case, the side jets
of the M-shaped velocity profile are caused by a geometrical
heterogeneity imposed by the sidewalls of the duct, so the
stagnant region tends to spread between the sidewalls. In the
latter case, maxima of streamwise velocity appear in an originally
free flow around the region where the magnetic field is of highest
intensity, and the core roughly corresponds to the region where
the magnetic field is imposed.

To our knowledge, most of numerical studies of fringing magnetic
fields were performed with the 2D assumptions, i.e. the flow was
treated as quasi 2D, where only the transverse field component
($B_z$) was taken into consideration, while the other components
($B_x$ and $B_y$) were neglected. As a result, the studied
magnetic field was inconsistent, that is, the requirements for the
field to be curl- and divergence-free were violated
\cite{Votyakov:TCFD:2009}. So, this is one of the contributions of
the present work: a complete systematic 3D numerical study with
$N$ changing smoothly from low to high, while maintaining a
physically consistent curl- and divergence-free external magnetic
field.

A new physical effect compared to the fringing magnetic fields is
a neat demonstration of the vortices alongside the stagnant core.
It has been shown recently that the spanwise homogeneous fringing
magnetic field does not enable any recirculation
\cite{Votyakov:JFM:2007}.

The second aspect of the paper is the detachment of the vortices
from the magnetic obstacle when the Reynolds number $Re$ is large
enough. Magnetic and connecting vortices are in rest during the
vortex shedding. The shedding can be either symmetric, in which
both attached vortices are coming off simultaneously, or
asymmetric, as it usually happens with a solid obstacle when $Re$
exceeds a critical value. The symmetric vortex shedding is also
possible in an ordinary hydrodynamic flow past an infinitely long
cylinder that is specially perturbed to provoke the synchronous
vortex shedding, see e.g. \cite{Konstantinidis:Balabani:2007}.

The presented results are  complementary to those published in
\cite{Votyakov:PRL:2007, Votyakov:JFM:2007}. They were unavailable
before because there required extensive sets of 3D numerical
simulations: a series of runs for large $N$ to refine the core of
the magnetic obstacle, and a series of runs involving long time
integrations to produce laminar time-periodic vortex shedding at
high $Re$. Each of these sets of runs is discussed below in its
own section.

The structure of the present paper is as follows. First, we
present technical details of the simulations: model, equations and
3D numerical solver. Then, we report results for the core of the
magnetic obstacle and demonstrate possible symmetric and
asymmetric vortex shedding in the wake past the obstacle. The last
Section before the Summary explains the generic scenario for the wake
of the magnetic obstacle and the similarities with the vortex
shedding past the solid obstacle. A summary of the main
conclusions ends the paper.

\section*{Model, equations, numerical method}\label{sec:problem}

A schematic of the model is shown in
Fig.~\ref{Fig:Introductory}($c$). It is the same as detailed in
\cite{Votyakov:JFM:2007} except for the fact that, in the present
case, we have no side walls. Instead, here we use slip boundary
conditions in the crosswise direction, and therefore, expand the
crosswise dimension of the computational domain. Also, in vortex
shedding simulations, we double the outlet length compared to
\cite{Votyakov:JFM:2007}   in order to exclude the outlet
influence on the vortex detachment and advection.

The governing equations for an electrically conducting and
incompressible fluid, subject to an external magnetic field, are
the Navier-Stokes equations coupled with the Maxwell equations for
a moving medium and Ohm's law. Here, the magnetic Reynolds number
$R_m=\mu^*{\sigma}u_{0}H$  is supposed to be much less than one,
where, $\mu^*$ is the magnetic permeability. This corresponds to
the so called quasi-static (or inductionless) approximation, where
it is assumed that the induced magnetic field is infinitely small
in comparison to the external magnetic field, see, e.g.
\cite{Roberts:1967}. The resulting equations in dimensionless form
are:
    \begin{eqnarray}
    \label{eq:NSE:momentum}
    \frac{{\partial\textbf{u}}}{{\partial t}} + (\textbf{u} \cdot \nabla ) \textbf{u}
    &=& - \nabla p + \frac{1}{Re}\triangle \textbf{u} + N
    (\textbf{j}\times\textbf{B}),
    \label{eq:NSE:continuity} \;\;\; \nabla \cdot \textbf{u} = 0,
    \\
    \label{eq:NSE:Ohm} \textbf{j} &=&   -\nabla\phi + \textbf{u} \times
    \textbf{B},
    \label{eq:NSE:Poisson} \;\;\;\; \nabla \cdot \textbf{j} = 0,
    \end{eqnarray}
where $\bf{u}$ denotes velocity field, $\bf{B}$ is the external
magnetic field, $\bf{j}$ is the electric current density, $p$ is
the pressure, and $\phi$ is the electric potential. The
interaction parameter $N$ and Reynolds number $Re$, $N=Ha^2/Re$,
are linked by means of the Hartmann number:
$Ha=HB_0(\sigma/\rho\nu)^{1/2}$. The Hartmann number determines
the thickness of the Hartmann boundary layers, $\delta/H \sim
Ha^{-1}$ for flow under constant magnetic field.

The origin of the coordinate system is taken in the center of the
magnetic gap. The computational domain is: $-L_{x1}\le x \le
L_{x2}$, $-L_{y}\le y \le L_{y}$, $-H \le z \le H$, where $x, y, z$
are respectively the streamwise, crosswise, and transverse
directions, and $L_{x1}$ ($L_{x2}$), $L_{y}$, and $H$ are the inlet
(outlet), crosswise, and transverse dimensions of the simulation
box. $L_{x2}=25$ in runs for the core of the magnetic obstacle, and
$L_{x2}=50$ in runs for vortex shedding; $L_{x1}=25$, $L_{y}=25$,
$H=1$ in both simulations. Magnetic poles are located at $x=0$,
$y=0$, $z=\pm h$, and the size of the magnet is $|x|\le M_x$,
$|y|\le M_y$, $|z|\ge h$. The intensity of the external magnetic
field ${\bf B(r)}$ is calculated by means of formulae given in
\cite{Votyakov:JFM:2007} with $M_x=1.5$, $M_y=2$, and $h=1.5$.
Different cuts of the intensity ${\bf B(r)}$ for these parameters
are plotted in Fig.~3 and Fig.~4($b$) in paper
\cite{Votyakov:JFM:2007}.

The characteristic dimensions for the Reynolds number $Re$, and
the interaction parameter $N$ are the half-height of the duct $H$,
the mean flow rate $u_0$, and the magnetic field intensity $B_0$,
taken at the center of the magnetic gap, $x\!=\!y\!=\!z\!=\!0$.
\bchange{So, all the distances $L_x, L_{x1}, L_{x2}, L_y, M_y,
M_x, h$ are normalized by $H$;  the velocity $\mathbf{u}$ by
$u_0$; the magnetic field $\mathbf{B}$ by $B_0$; the electric
current density $\mathbf{j}$ by $\sigma u_0 B_0$; the electric
potential by $H u_0 B_0$; the pressure $p$ by $\rho\nu u_0/H$.}

For a given external field ${\bf B}(x,y,z)$, the unknowns of the
partial differential equations (\ref{eq:NSE:momentum} --
\ref{eq:NSE:Poisson}) are the velocity vector field ${\bf
u}(x,y,z)$, and two scalar fields: the pressure $p(x,y,z)$ and the
electric potential $\phi(x,y,z)$. To find the unknowns we use a
finite differences method that was implemented in a 3D numerical
solver as been detailed in\cite{Votyakov:Zienicke:FDMP:2006}. The
solver was developed from a free hydrodynamic solver created
originally in the research group of Prof.~M.~Griebel
(\cite{Griebel:book:1995}). The solver employs the Chorin-type
projection algorithm and finite differences on an inhomogeneous
staggered regular grid. Time integration is done by the explicit
Adams-Bashforth method that has second order accuracy. Convective
and diffusive terms are implemented by means of the VONOS
(variable-order non-oscillatory scheme) method. The 3D Poisson
equations are solved for pressure and electric potential at each
time step by using the bi-conjugate gradient stabilized method
(BiCGStab).

To complete the numerical model, boundary conditions have to be
specified. No slip and insulating walls were specified in the
transverse direction, while slip walls were used in the crosswise
direction. In order to test the effect of boundary conditions, in
some of the runs carried out for the core of the magnetic
obstacle, the slip conditions in the crosswise direction were
replaced by periodic boundary conditions. However, changing the
crosswise boundary conditions was found to have no effect on the
structure of the core. This is because $L_y=25$ is large enough
compared to $M_y=2$ in all the runs for the core.

The outlet of the computational domain was treated as a force-free
(straight-out) border for the velocity. The electric potential at
the inlet and outlet boundaries was taken to be equal to zero
because the inlet and outlet are sufficiently far from the region
of magnetic field. At the inlet, a 2D parabolic (Poiseuille)
velocity profile was used that was uniform in the crossflow
(spanwise) direction. In runs for the asymmetric vortex shedding,
this profile was slightly perturbed in the crosswise direction at
initial times and then kept symmetric and constant in time. The
initial perturbation initiated the asymmetric vortex shedding and
the followed symmetry and constancy assured that the asymmetric
vortex shedding is independent of the inlet conditions.

Time integration in the runs for the core of the magnetic
obstacles was carried out until a stationary laminar solution has
been reached. In all these simulations, we found the same laminar
solution at a given $Re$ and $N$ pair, independently of initial
conditions. So, as initial conditions for runs corresponding to
new $Re$ and $N$ we used 3D fields of velocity, pressure, and
electric potential obtained from the previous runs having the
closest $Re$ and $N$ values.

Time integration in the runs for the vortex shedding was continued
until a time-periodic laminar solution was reached. These
simulations were dependent on the initial conditions. We found two
classes of solutions: symmetric and asymmetric distribution of
attached vortices in the wake at large times. The details about
the initial and inlet conditions for both cases are given in the
beginning of the corresponding sections.

The simulation box has been discretized by an inhomogeneous
regular 3D grid  depending on the solved problem. Details about
the numerical grid are given at the beginning of the corresponding
sections.

\section*{Core of the magnetic obstacle} \label{sec:core}
In this series of simulations, we focus on the flow around a
magnetic obstacle at large interaction parameter $N$. In order to
achieve large $N=Ha^2/Re$, the simulations were started at a small
interaction parameter and $Ha$ was smoothly increased, while
keeping $Re$ constant. Two values of the Reynolds number were
studied, $Re=10 \mbox{ and } 100$, and no principal differences
were found at the same $N$. These low values of $Re$ imply low
inertial forces, therefore, only ``two-vortex'' patterns were produced,
without connecting or attached vortices.

The numerical grid was regular and inhomogeneous, $N_x\times N_y
\times N_z=64^3$. The minimal horizontal step size in the region of
the magnetic gap was $\Delta x \simeq \Delta y \simeq 0.3$, which
means that a few dozens points were used for resolving the inner
vortices in the core of the magnetic obstacle. The minimal vertical
step size near the top and bottom (Hartmann) walls was $\Delta
z=0.005$. This corresponds to using three to five ($=(1/Ha)/\Delta
z)$ points to resolve Hartmann layer at $Ha=40-70$.

\begin{figure}
\begin{center}
    \includegraphics[width=14cm, angle=0, clip=on]{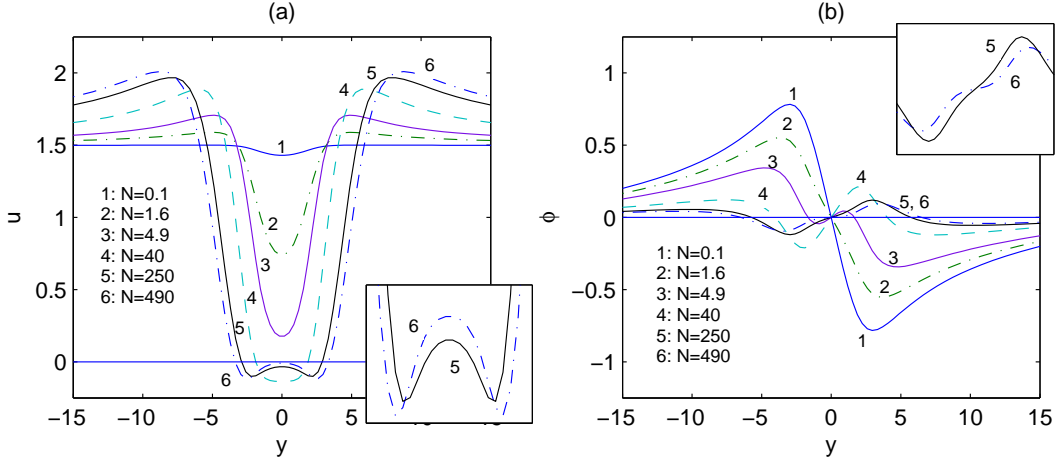}
\end{center}
    \caption{\label{Fig:u_phi_dist}Streamwise velocity ($a$) and electric
    potential ($b$) along crosswise cuts of middle horizontal plane
    $x=z=0$. $Re=10$, $N=$0.1(solid 1), 1.6(dot-dashed 2), 4.9(solid 3),
    40(dashed 4), 250(solid 5), and 490(dot-dashed 6). Insertion shows
    magnified plots for curves 5 and 6.}
\end{figure}

The easiest way to understand the core of the magnetic obstacle is
to analyze crosswise cuts through the center of the magnetic gap at
different arising magnetic interaction parameters $N$. These cuts
are shown in Fig.~\ref{Fig:u_phi_dist}a for the streamwise velocity
$u_x(y)$ and in  Fig.~\ref{Fig:u_phi_dist}b for the electric
potential $\phi(y)$. First we discuss how the streamwise velocity
changes as  $N$ increases.

Because $N$ expresses the strength of the retarding Lorentz force
relative to the inertial force, curve 1 in
Fig.~\ref{Fig:u_phi_dist}$a$ ($N=0.1$) is only slightly disturbed
with respect to a constant. As $N$ increases, the curves $u_x(y)$
pull further down in the central part $u_{center} \doteqdot
u_x(0)$, see for example curves 2 and 3. At $N$ higher than a
critical value $N_{c,m}$, i.e. for curve 4, the central velocities
$u_{center}$ are negative. This means that there appears a reverse
flow causing magnetic vortices in the magnetic gap. When $N$ rises
even more (see curves 5 and 6) the magnetic vortices become
stronger and simultaneously shift away from the center to the side
along the $y$ direction, see insertion in
Fig.~\ref{Fig:u_phi_dist}$a$ for curves 5 and 6.

The fact that the centerline velocity in the center of the
magnetic gap goes to zero as $N$ increases is expected and was
discussed earlier for fringing magnetic fields. In this respect,
the case of the magnetic obstacle is analogous to the fringing
magnetic field. What is different in these two cases is that the
centerline velocity becomes negative before it goes to zero while
this could not be so, and was never actually observed, for the
fringing field \cite{Votyakov:JFM:2007}.

Fig.~\ref{Fig:u_phi_dist}($b$) shows how the electric potential
$\phi(y)$ varies along the central crosswise cut through the
magnetic gap. The slope in the central point is the crosswise
electric field, $E_{y,center}=-d\phi/dy|_{y=0}$. One can see that
$E_{y,center}$ changes its sign:  it is positive at small $N$ and
negative at high $N$. To explain why it is so, one can use the
following way of thinking. Any free flow tends to pass over an
obstacle in such a way so as to perform the lowest possible
mechanical work, i.e. flow streamlines are the lines of least
resistance to the transfer of mass. The resistance of the flow
subject to an external magnetic field is caused by the retarding
Lorentz force $F_x\approx j_y B_z$, so the flow tends to produce a
crosswise electric current, $j_y$, as low as possible while
preserving the divergence-free condition $\nabla\cdot{\bf j}=0$. To
satisfy the latter requirement, an electric field ${\bf E}$ must
appear, which is directed in such a way, so as to compensate the
currents produced by the electromotive force ${\bf u}\times{\bf
B}$. Next, we analyze the crosswise electric current $j_y = E_y +
(u_z B_x - u_x B_z)$. Due to symmetry in the center of the magnetic
gap $B_y=B_x=u_y=u_z=j_y=j_z=0$  so $j_y = E_y - u_x B_z$. This
means that  $E_y$ tends to have the same sign as $u_x$ in order to
make $j_y$ smaller. At small $N$, the streamwise velocity $u_x$ is
large and positive, so the electric field $E_y$ is positive too.
When the magnetic vortices appear, there is a reverse flow in the
center. Therefore, the central velocity is negative now, and the
central electric field $E_{y,center}$ is also negative.

In \cite{Votyakov:JFM:2007} the change of the electric field in
the magnetic gap is explained in terms of the Poisson equation and
the concurrence between external and internal vorticity. This
argument is also valid here, however in contrast to
\cite{Votyakov:JFM:2007}, we have no side walls,  so the external
vorticity in the present case plays only a minor role. As a
result, the reversal of the electric field appears at a small $N$
(approximately equal to five), \bchange{which is close to the
critical interaction parameter $N$ at $\kappa=0.4$ given in
\cite{Votyakov:JFM:2007}}. (In \cite{Votyakov:JFM:2007}, $\kappa$
is the ratio of the magnet width to the duct width.)

\begin{figure}
\begin{center}
    \includegraphics[width=14cm, angle=0, clip=yes]{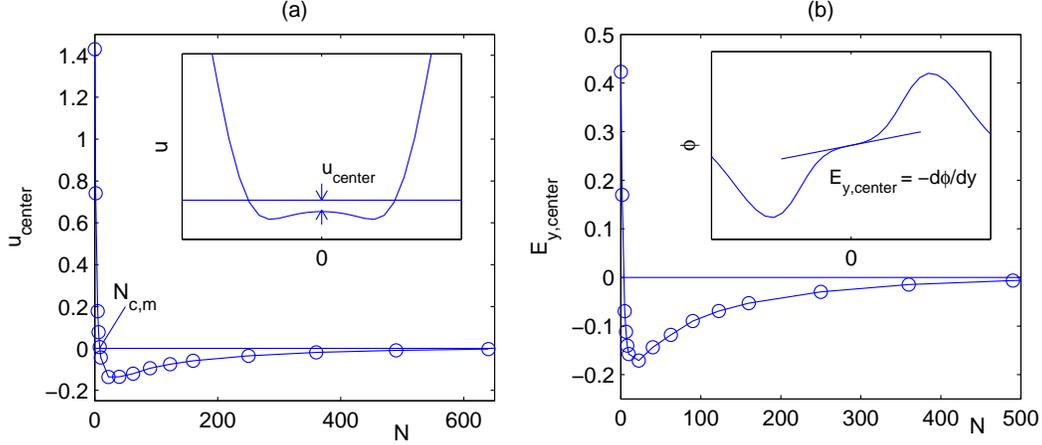}
\end{center}
    \caption{\label{Fig:u_phi_summ}Central streamwise velocity $u_{center}$ ($a$)
    and central spanwise electric field $E_{y,center}$ ($b$) as a function
    of the interaction parameter $N$. $N_{c,m}$ is a critical
    value where the streamwise velocity is equal to zero.
    Insertion shows the definition of $u_{center}$ and $E_{y,center}$.}
\end{figure}

The overall data about $u_{center}$ and $E_{y,center}$ in the
whole range of studied $N$ are shown in Fig.~\ref{Fig:u_phi_summ}.
One can see that both characteristics start from positive values,
then, they cross the zeroth level, reach a minimum, go up again,
and finally vanish in the limit of high $N$. With respect to the
streamwise velocity, this means that, at hight $N$, there is no
mass flow in the center of the magnetic gap; the other velocity
components are equal to zero due to symmetry. With respect to the
crosswise electric field, this means that there are no electric
currents. This occurs because there is no mass flow, therefore,
the electromotive force vanishes, $E_y$ goes to zero, and the
other electric field components are equal to zero due to symmetry.
Thus, one can say that the center of the magnetic gap is frozen by
the strong external magnetic field, so that both mass flow and
electric currents tend to bypass the center. In other words, this
means that a strong magnetic obstacle has a core, and such a core
is like a solid insulated body, being impenetrable for the
external mass and electric charge flow.

\bchange{In the 2D creeping flow around the magnetic obstacle, the
$u_{center}$ and $E_{y,center}$ vanishing  behavior shown in
Fig.~\ref{Fig:u_phi_summ} is impossible because it violates the flow
continuity. At very high $N$ and low $Re$, instead of the frozen core
obtained in the 3D case, a 2D flow develops various
recirculation patterns in the core, because the secondary flow of the 3D
magnetic vortices is forbidden in the 2D case. Paradoxically, the
2D creeping flow discussed in the paper by Cuevas \textit{et al.}
\cite{Cuevas:Smolentsev:Abdou:PRE:2006} is turned out to be more
rich than the presented 3D creeping flow between two no-slip
endplates. This point is discussed further in the last Section
devoted to the generic scenario of the wake of the magnetic
obstacle.}

\begin{figure}
\begin{center}
    \includegraphics[width=13.5cm, angle=0, clip=yes]{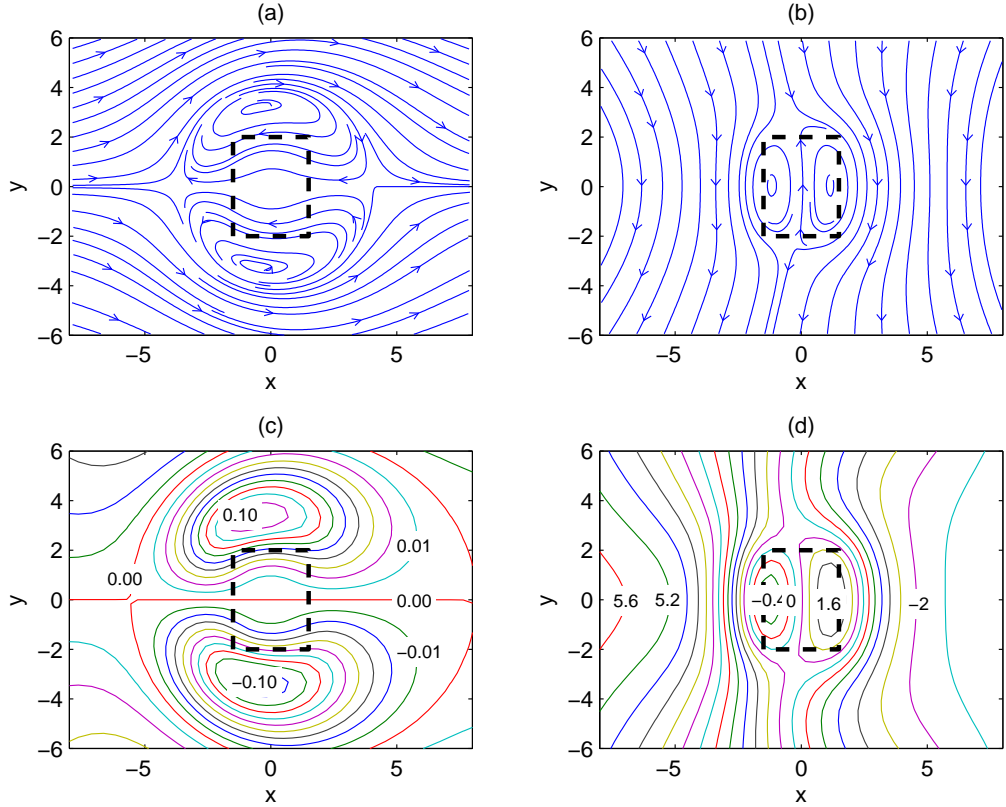}
\end{center}
    \caption{\label{Fig:phi_p_u_jRe0010Ha070}
    Middle horizontal plane, $z=0$:
    streamlines of the mass ($u_x, u_y$) ($a$)  and electric charge
    ($j_x, y_y$) ($b$) flow. Contour lines for the electric potential $\phi(x,y)$ ($c$)
    and pressure $p(x,y)$ ($d$) resemble the streamlines given above.
    $Re=10$,    $N=490$. Contours of the electric potential are given with step 0.01,
    and contours of the pressure are given with the step 0.4.
    Dashed bold rectangle shows borders of the external magnet.}
\end{figure}

It is convenient to visualize the core of the magnetic obstacle by
plotting streamlines for the mass flow, (see
Fig.~\ref{Fig:phi_p_u_jRe0010Ha070}$a$) and electric charge
transfer (see Fig.~\ref{Fig:phi_p_u_jRe0010Ha070}$b$) in the
middle horizontal plane. One can see that the side streamlines
envelop the bold dashed rectangle. This rectangle denotes the
borders of the external magnet. Alongside the rectangle there are
closed streamlines for mass flow (plot $a$), which are magnetic
vortices. At high $N$, these vortices are located in the region of
crosswise gradients of the external magnetic field and compensate
shear stresses between the core of the magnetic obstacle and rest
of the flow. Also, the magnetic vortices produce closed electric
currents inside the rectangle (plot $b$). These internal currents
are elongated in the $y$ direction. They are very weak compared to
the external currents enveloping the obstacle.

We note that the streamlines of the flow and electric charge
resemble contour lines of the electric potential
(Fig.~\ref{Fig:phi_p_u_jRe0010Ha070}$c$) and pressure
(Fig.~\ref{Fig:phi_p_u_jRe0010Ha070}$d$). This happens because
inertia and viscosity are vanishing in the core, so equations
(\ref{eq:NSE:momentum} -- \ref{eq:NSE:Poisson}) become:
$$
\nabla p = {\bf j \times B}, \quad \nabla \phi = -{\bf j}+{\bf
u\times B} \approx {\bf u\times B}\,.
$$
In the latter equation, ${\bf j} \ll \nabla\phi \mbox{ and } {\bf
u\times B}$ is the dominating term. In the core of the obstacle
${\bf B}=(0,0,B_z)\approx (0,0,1)$, hence, the pressure (electric
potential) is a streamline function for the electric current
(velocity). These relationships for the region of the flow subject
to the strong magnetic field had been discussed earlier by
Kulikovskii in 1968 \cite{Kulikovskii:1968}.

Kulikovskii's theory is linear, therefore, it must work well in
the stagnant core of the magnetic obstacle. The traditional
approach in the limits of this theory is to introduce  so-called
characteristic surfaces and then to impose Hartmann layers as
boundary conditions for further integration along the
characteristic surfaces. Such an approach has been used before for
slowly varying fringing magnetic fields \cite{Alboussiere:2004},
where  Hartmann layers and inertialess assumption are reasonable.
However, it is an open issue whether the conception of the
characteristic surfaces is valid for the case of the magnetic
obstacle. For perfectly electrically conductive liquids this
conception enforces mass and electric streamlines to flow along
the surfaces of constant $\mathbf{B}$ what obviously not the case
shown in Fig.~\ref{Fig:phi_p_u_jRe0010Ha070}$a,b$.

Magnetic field plus rotation require more sophisticated boundary
conditions than just the Hartmann layer. There is known a solution
for the Ekman-Hartman layers, where both constant rotation and
constant magnetic field are taken jointly into account. This
probably does not also fit  because the vorticity is not constant
along the transverse direction, and the shape of vortices is not
circular. Moreover, inclusion of the non constant vorticity
destroys the linearity of Kulikovskii's theory. Therefore,
Kulikovskii's theory could not be used as it stands to predict
recirculation \emph{a priori}. Indeed this explains why the theory
has not been applied to magnetic vortices, even though it has been
known for a while. Nevertheless, Kulikovskii's theory is useful
and must be mentioned because it explains \emph{a posteriori} the
shape of vortices and their matching to electric potential lines.

\section*{Vortex shedding past a magnetic obstacle} \label{sec:shedding}
The following simulations were carried out at $Re=900$ and $N=9$
($Ha=90$). The numerical grid is regular and inhomogeneous,
$N_x\times N_y \times N_z=144\times 96 \times 64$. The minimal
horizontal step size is $\Delta x \simeq \Delta y \simeq 0.25-0.33$
in the region $|x|\le 3M_x$, $|y|\le 2M_y$. This supplies few
dozens points inside and near the magnetic gap, enough to resolve
recirculation. The vertical step size near the top and bottom
(Hartmann) walls is $\Delta z=0.009$. Otherwise, it is impossible,
at the computational power existing nowadays, to perform lengthy
time dependent 3D simulations in a box that is long enough to
observe a vortex shedding. We believe that this vertical step size
is sufficient because of the following reasons. The magnetic field
decays quickly, therefore the thickness of boundary layers quickly
increases. Moreover, the region of highest magnetic intensity is in
the center of the core of the magnetic obstacle and is
characterized by low velocities. Finally, it has been shown that
ignoring the Hartmann friction destabilizes the magnetic and
connecting vortices \cite{Votyakov:JFM:2007}. In our simulations,
these vortices remained stable during all simulations, therefore,
we believe that the vertical resolution was enough for the purpose
of the paper under consideration.

All the results below are shown for the mid central plane, where
all vortex peculiarities can be distinctively visualized.
Nevertheless, it is necessary to note that the flow in the mid
plane is not two-dimensional. There is a secondary flow from and
into the mid plane towards and from the top and bottom walls. This
secondary flow is caused by the process of creation and
destruction of the Hartmann layers. 3D pictures of the vortices
have been drawn before \cite{Votyakov:JFM:2007} and will not be
considered here. Shortly, 3D peculiarities are that the vortices
are stabilized by friction with the top and bottom walls. For the
magnetic vortices, this is the Hartmann friction, and for the
attached vortices, this is the viscous friction. The mission of
the connecting vortices is to make consistent the rotation of the
magnetic and attached vortices, therefore, the connecting vortices
are retained by the magnetic and attached vortices. As a result,
the connecting vortices  are stabilized jointly by both the
Hartmann friction and viscous friction.

\begin{figure}
\begin{center}
    \includegraphics[width=13cm, angle=0, clip=yes]{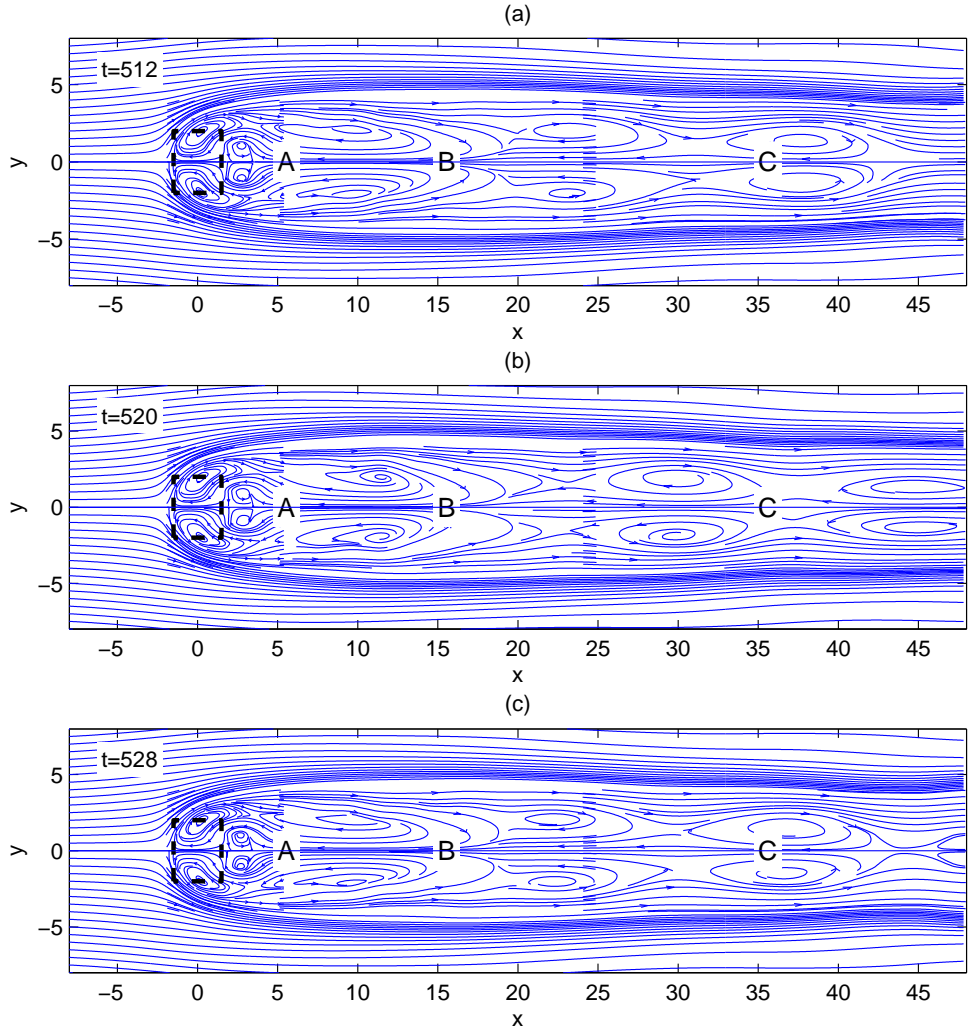}
\end{center}
    \caption{\label{Fig:symmshedding}
    Instantaneous mass flow streamlines for the unperturbed symmetric inlet velocity
    profile: $t=$512($a$), 520($b$), 528($c$). $Re=900$, $N=9$.
    Dashed bold rectangle shows borders of the external magnet.
    Letters A at $x=5$, B at $x=15$, and $C$ at $x=35$ are points on
    the centerline, $y=0$, for time histories shown
    as dashed lines in Fig.~\ref{Fig:dpdxdpdy}.}
\end{figure}

\begin{figure}
\begin{center}
    \includegraphics[width=13cm, angle=0, clip=yes]{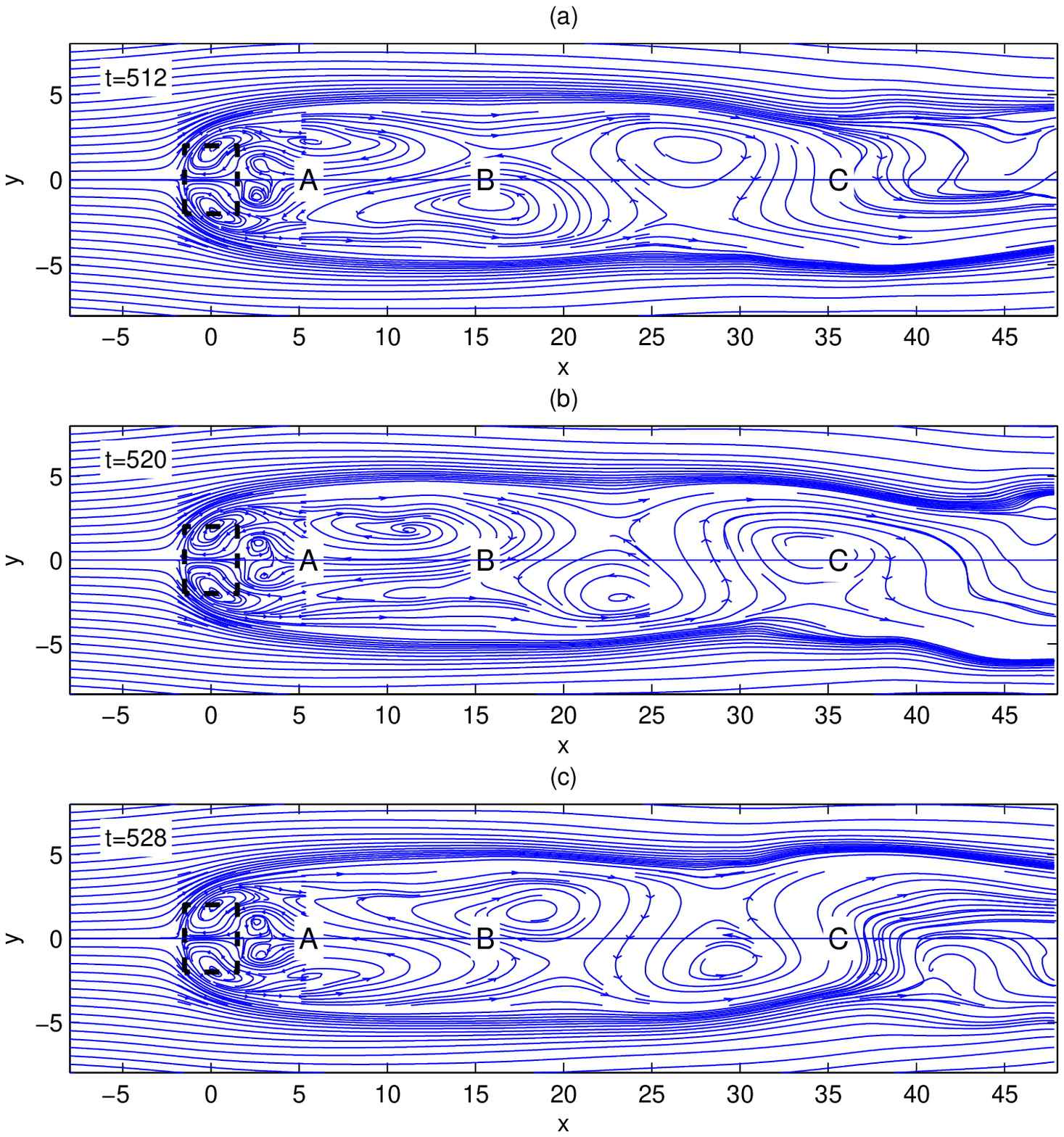}
\end{center}
    \caption{\label{Fig:asymmshedding}
    Instantaneous mass flow streamlines for the initially
    perturbed symmetric inlet velocity
    profile: $t=$512($a$), 520($b$), 528($c$). $Re=900$, $N=9$.
    Dashed bold rectangle shows borders of the external magnet.
    Letters A at $x=5$, B at $x=15$, and $C$ at $x=35$ are points on
    the centerline, $y=0$, for time histories shown
    as solid lines in Fig.~\ref{Fig:dpdxdpdy}.}
\end{figure}

We studied two classes of initial and inlet conditions:
unperturbed and perturbed at time $0\le t \le 120$. The former
resulted in the time periodic symmetric vortex shedding
(Fig.~\ref{Fig:symmshedding}) and the latter resulted in the
standard asymmetric vortex shedding
(Fig.~\ref{Fig:asymmshedding}). Crosswise and transverse velocity
components were taken equal to zero at $t=0$ in both cases.
The integration time step varied automatically with the limitation
imposed by the viscous layer stability condition. The largest time
step was equal to 0.0083.

For the unperturbed case, we used the initial ($t=0$) streamwise
component velocity $u_{x,s}(x,y,z) = u_P(z)$, where the
$u_P(z)=3/2(1-z^2)$ is the Poiseuille velocity profile. The same
inlet velocity profile $u_{x,s}(0,y,z)$ was imposed at all times
and this explains why we called this case unperturbed. Time
integration was stopped at $t=612$.

For the perturbed case we used for the initial ($t=0$) streamwise
velocity $u_{x,a}(x,y,z)= u_P(z)\theta(y)$, where function
$\theta(y)=1$ for $|y|\ge \lambda$, and $\theta(y)=(1+\gamma
sin[\pi\frac{y}{\lambda}])$ for $|y|\le \lambda$. The wavelength
of perturbation, $\lambda$, was taken to be sufficiently higher
than the spanwise size of the physical magnet, $\lambda=2.5M_y$.
The amplitude of the perturbation,  $\gamma$, was taken to be
$\gamma=0.05$. This five percent skew was sufficient to avoid the
symmetric solution found in the unperturbed case above. However,
it also resulted in slightly different flow rates for positive and
negative $y$. Therefore, the perturbed profile $u_{x,a}(0,y,z)$
was imposed at $0\le t \le 120$ only, and after $t=120$ the
symmetric profile $u_{x,s}(0,y,z)$ was prescribed again to restore
by that the equal flow rates.  Time integration was stopped at
$t=1100$.

Instantaneous mass flow streamlines for symmetric vortex shedding
are shown in Fig.~\ref{Fig:symmshedding}. They are plotted for the
middle horizontal plane, $z=0$, and are symmetric with respect to
the centerline $y=0$. There are three instances of time with
$\Delta t=8$. For each, we see the same configuration of magnetic
(first pair) and connecting (second pair) vortices, while attached
vortices form a sequence dependent on the time instance. This
provides evidence that the attached vortices come off the magnetic
obstacle simultaneously and move downstream slowly.  The location
of the attached vortices in plot~($a$) ($t=512$) looks similar as
in plot~($c$) ($t=528$), therefore one can conclude that vortex
breakdown occurs at a time period equal to 16 time units.

Although symmetric vortex shedding past a bluff body is not
typical in ordinary hydrodynamics, it is possible if one takes
special steps, such as artificial forcing, see for instance
\cite{Konstantinidis:Balabani:2007}, and references in Table 1
therein. The overwhelming majority of papers devoted to vortex
shedding deals with an infinitely long cylinder. The MHD case
under consideration is three-dimensional, i.e. there are top and
bottom walls, so the proper hydrodynamic analogy is to consider a
finite cylinder placed perpendicular between two endplates. There
is evidence in ordinary hydrodynamics that the confinement imposed
by the endplates increases the stability of the wake, see e.g.
\cite{Shair:Grove:etal:1963}, \cite{Nishioka:Sato:1974},
\cite{Gerich:Eckelmann:1982}, \cite{Lee:Budwig:1991}. In
particular, the range of the Reynolds number, $Re$, where two
attached vortices remain symmetric behind a circular cylinder
without breaking, is much larger in the presence of no-slip
endplates. Therefore, it is also possible that the confinement
stabilizes symmetric vortex shedding produced by the solid
cylinder subject to the artificial forcing.

If the inlet velocity profile is not symmetric at initial times,
then one expects an asymmetric vortex shedding, as shown in
Fig.~\ref{Fig:asymmshedding}. The time instances are the same as
in Fig.~\ref{Fig:symmshedding}. One can see now that the attached
vortices are shifted  relative to each other through the
centerline $y=0$. Plot~$a$ ($t=512$) looks roughly like the mirror
image of plot~$c$ ($t=528$) giving by that evidence about the
half-time period equal approximately to 16 time units. Altogether,
the picture is similar to a standard time-periodic laminar vortex
shedding past a solid circular cylinder. At the place of the
cylinder there is a four-vortex ensemble composed of magnetic and
connecting vortices. Because of Hartmann friction, this ensemble
is stable in time, and so this represents the body of a virtual
\bchange{solid} obstacle imposed by the external, strongly
heterogeneous magnetic field.

\begin{figure}
\begin{center}
    \includegraphics[width=13.5cm, angle=0, clip=yes]{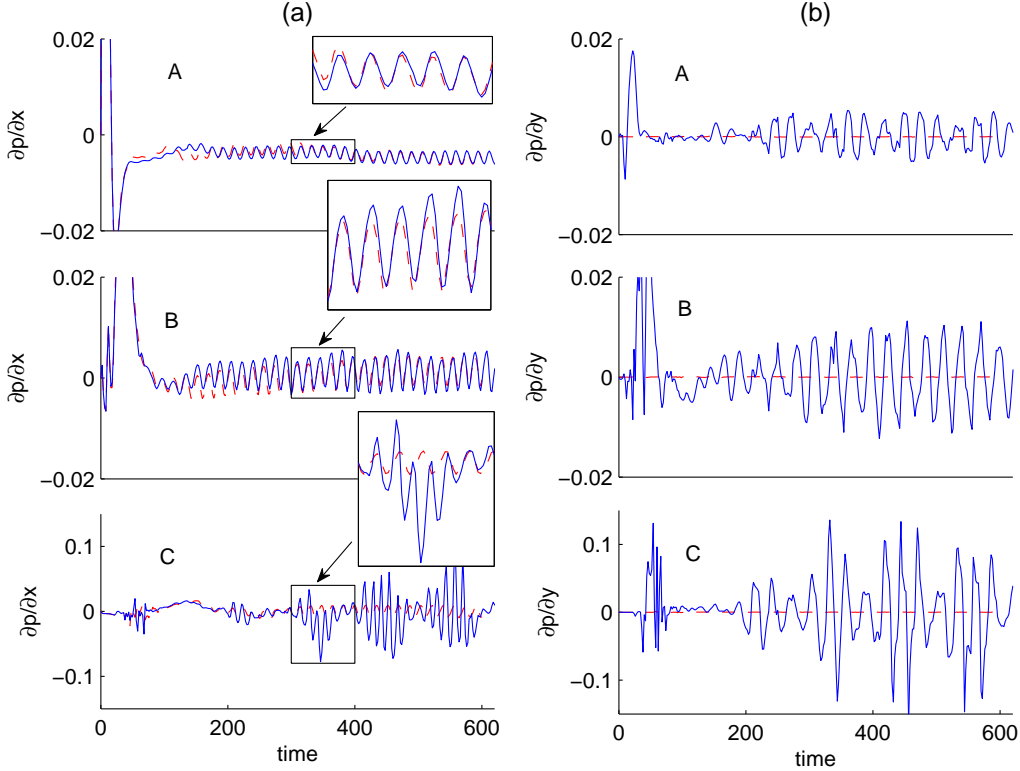}
\end{center}
    \caption{\label{Fig:dpdxdpdy}
    Streamwise ($\partial p/\partial x$, plot $a$)
    and crosswise ($\partial p/\partial y$, plot $b$) pressure
    gradients in points $A$, $B$, $C$ given in
    Fig.~\ref{Fig:symmshedding} (dashed) and \ref{Fig:asymmshedding} (solid lines).
    Insertions are zoomed for the time range $300\le t \le 400$.}
\end{figure}

Shown in Fig.~\ref{Fig:dpdxdpdy} are time histories of local
instantaneous streamwise ($\partial p/\partial x$, plot $a$) and
crosswise ($\partial p/\partial y$, plot $b$) pressure gradients,
for both symmetric (dashed) and asymmetric (solid lines) vortex
shedding. The curves are given for points $A,B,C$ located on the
centerline $y=0$ and denoted in Fig.~\ref{Fig:symmshedding} and
Fig.~\ref{Fig:asymmshedding}. These pressure gradients are selected
for time analysis because they can be measured experimentally.
Moreover,  $\partial p/\partial x$ and $\partial p/\partial y$ can
be understood as the local drag and lift forces respectively. Time
dependencies of drag and lift coefficients in the case of a solid
cylinder are well understood, see e.g. Fig.~7 in
\cite{Singh:Mittal:2005}. They are time-periodic with a single
vortex shedding frequency at low Reynolds number.

As one can see in Fig.~\ref{Fig:dpdxdpdy}, the simulated $\partial
p/\partial x$, $\partial p/\partial y$ first go through a
transitional regime, which it is then transformed smoothly into a
periodic regime. The time period, equal approximately to 16,  can
be estimated from the zoomed insertions plotted for the time range
$300\le t \le 400$. (In these time dependent simulations the data
were recorded every two time units, so the precision of the time
period is plus/minus one.) \bchange{It is easy to estimate roughly
the Strouhal number $St=fL/u$. Here, $f=1/16$ is the frequency of
vortex shedding, $L=2M_y=4$ is the crossstream size of the magnet,
and $u=1$ is the mean flow velocity, so $St=1/4$. This value is different
from that of 0.1 found in the paper by Cuevas \textit{et al.} in 2D
simulations. The difference is obviously explained by the impact
of the channel walls.} For the symmetric (dashed) and asymmetric
(solid) vortex shedding, the streamwise pressure gradient has
similar behavior at locations not far from the magnetic obstacle.
One can see that $\partial p/\partial x$'s are very close at point
$A$, slightly disagree at point $B$ and notably different at point
$C$. The crosswise pressure gradient on the centerline for the
symmetric shedding is equal to zero.

\section*{The generic scenario for the wake of the magnetic
obstacle} \label{sec:scenario}

In ordinary hydrodynamics, the generic scenario for the wake
past the solid obstacle is the following when $Re$ smoothly
increases \cite{Feynman:Lectures:Vol2:1964} before turbulence
starts: (i) creeping flow, (ii) two attached vortices, (iii)
vortex shedding. If one goes into the details, e.g. considers the
different ways of vortex shedding, then different sub-scenarios
can be found depending on specific conditions,  but the generic
character  of the above classification remains. We emphasize that
the interplay between the viscous and inertial forces is decisive
for establishing this general peculiarities.

Analogously, by taking into consideration all possible forces, we
derive now a scenario for the wake of the magnetic obstacle.  This
was given briefly before in \cite{Votyakov:PRL:2007,
Votyakov:JFM:2007} without stressing that this is generic because
of the lack at the time of information about vortex shedding. Now,
this gap is filled.

In the case of the magnetic obstacle, there are three forces and
three corresponding terms in the MHD equations: the viscous force
(V), the inertial force (I), and the Lorentz force (L). If we put
the forces \bchange{in order of decreasing intensity}, then
the total number of all the possible relationships between forces
is six: (1)  VIL, (2) IVL, (3) VLI, (4) LVI, (5) LIV, and (6) ILV.
(Each capital letter is given for the corresponding term.) The
cases (1-2) are of the smallest Lorentz term, therefore, they can
be treated  as outlined above for the ordinary hydrodynamics. The
cases (3-4) are of the smallest inertial term, therefore, there
should be no attached vortices past the magnetic obstacle, so the
possible scenario are either no vortices  when the Lorentz force
is smaller than the viscous force (case 3) or two alongside
magnetic vortices when the Lorentz force is larger than the
viscous force (case 4). Finally, the cases (5-6) are of the
smallest viscous term and the peculiar patterns are either six
vortices (case 5) when the inertial force is so low that the
attached vortices are retained or vortex shedding with specific
four-vortex pattern (case 6) as shown in previous Section. In the
latter, the four vortices taken together is an analog of the bluff
body as in the ordinary hydrodynamics.

\bchange{It is important to stress that we discuss a 3D flow
between two horizontal no-slip endplates. This discussion might be
projected onto a 2D flow, but carefully. For instance, the 2D flow
could not produce a 2D region excluded from the flow without
violating the continuity requirement,
$\nabla_\perp\mathbf{u}_\perp=0$. In a 3D flow, the latter
equation is
$\nabla\mathbf{u}=\nabla_\perp\mathbf{u}_\perp+\partial_{z}u_z=0$,
which can be satisfied by the secondary flow in the third
direction, i.e. by the  $\partial_{z}u_z$ term. This results in
the helical streamlines of magnetic vortices, see Fig.~11 in the
paper by Votyakov \textit{et al.} \cite{Votyakov:JFM:2007}.
However, a 3D helix could not be realized in a 2D space, so the
$u_{center}$, $E_{y,center}$ vanishing behavior shown in
Fig.~\ref{Fig:u_phi_summ} becomes impossible. Instead, in the
creeping 2D flow, $u_{center}$ is decreases as $N$ increases.
Then, at some high critical $N$, a $u_{center}$ drop does stop and
makes a flip into positive values causing two additional vortices
in the core of the obstacle. This new resulting flow structure
consists of four vortices as shown in Fig.~9$b$ of the paper by
Cuevas \textit{et al.} \cite{Cuevas:Smolentsev:Abdou:PRE:2006}. If
$N$ increases further, then even more intricate recirculation
patterns  are produced\footnote{\bchange{We performed 2D
simulations and found also four vortices shown  in the paper by
Cuevas \textit{et al.} \cite{Cuevas:Smolentsev:Abdou:PRE:2006}.
Moreover, other vortex configurations not reported in
\cite{Cuevas:Smolentsev:Abdou:PRE:2006} have been revealed at
higher $N$. The results will be submitted.}}. It looks paradoxical
that a 3D flow has the simpler structure than a 2D flow, however,
such simpler behavior is governed by a strong magnetic field,
prohibiting a penetration into the core, and by the secondary flow
in a vertical direction towards to Hartmann layers. Moreover, it
is an issue how to practically realize a 2D creeping flow in order
to reveal intricate recirculations in the 2D core without impact
of top and bottom flow boundaries.}

\bchange{Another} question that arises is whether the core of the magnetic
obstacle appears for a sufficiently high $Re$ even at high $N$,
that is, whether the upstream flow penetrates the stagnant core
made of the magnetic and connecting vortices. \bchange{Again}, the
answer depends on whether the flow considered is two- or
three-dimensional. We suggest that the magnetic vortices must
appear in both cases because $N$ is supposed to be sufficiently
high to produce recirculation. So the question can be
reformulated: whether the vortices and the stagnant core are
stable.

In 2D simulations there is no sink for the upstream kinetic energy
accumulated by the magnetic vortices. As a result, the rotating
magnetic vortices are not fixed in their location and move freely
in the plane  by responding to the pulsations of the upstream flow.
This destroys the core of the magnetic obstacle so it becomes
penetrable. If there are small time-dependent pulsations in the
upstream flow, then one can observe different (even exotic)
configurations of magnetic vortices, which can be mistakenly taken
as sub-scenarios of the given 2D simulation.

For any 2D approach applied to a realistic system, the main problem
is whether 2D assumptions are reasonable because the realistic
system has always endplates to hold magnetic poles. That is, the
Hartmann friction and viscous friction are always present. Of
course, to make 2D middle plane flow, the magnetic poles can be
moved far apart while synchronously increasing the magnetic field
intensity $B_0$ to keep the same $Ha$. But then, the gradient of
the magnetic field becomes more smooth, therefore, one needs a
higher critical  $N_c$ to observe recirculation
\cite{Votyakov:JFM:2007}.

In 3D simulations, there is a sink for the kinetic energy because
the mass streamlines, that form magnetic vortices, represent
helical trajectories into the Hartmann layers as shown recently in
\cite{Votyakov:JFM:2007}. Then, the pulsations of the upstream
kinetic energy are dissipated in the Hartmann layers by means of
the friction with top/bottom walls. As a result, the rotating
magnetic vortices are well fixed in their location and they do not
move freely. Thus, when Hartmann layers are properly resolved in
the 3D simulations and $N$ is enough high to induce alongside
recirculation, then the core of the magnetic obstacle must be
visible even at high $Re$.

\bchange{To make more clear the aforesaid statement we consider
the following two examples. First, one imagines a flow of moderate
$Re_1$ and such a high $Ha_1$ that the core of the magnetic obstacle
is stable. Then, by keeping $Ha_1$ constant, one increases $Re$,
e.g. by taking a higher flow rate, to find $Re_c(Ha_1)$ where the
core destabilizes. This happens at a critical value
$N_{c,1}=Ha_1^2/Re_c$. Now, one imagines a flow at moderate $Ha_2$
and such a high $Re_2$ that the core of the magnetic obstacle does not
exist. Then, while keeping $Re_2$ constant, one increases $Ha$, e.g.
by imposing a higher external magnetic field. There exists such a high
$Ha_c(Re_2)$ where the core stabilizes again. This happens at a
critical value $N_{c,2}=Ha_c^2/Re_2$, which is supposed to be of
the same order of magnitude as $N_{c,1}$. In other words, at any
high constant $Ha$ it is possible to find $Re$ destabilizing the
core, and vice verse, at any high constant $Re$, it is possible to
find $Ha$ stabilizing the core.}

Unfortunately, to confirm the above inference numerically is
impossible because it requires expensive 3D simulations where the
Hartmann layers must be properly resolved. For high $Re$ and $N$,
the thickness of the Hartmann layers is $1/\sqrt{Re*N}$. Then, e.g.
for $N=100$ (to guarantee magnetic vortices)  and big $Re$ the grid
resolution must be around $\delta/(10\sqrt{Re})$, where
$1/\delta=3...10 $  is the number of points to resolve Hartmann
layers. If $Re=10000$, then around $(L_y/H) (L_x/H) 10^{12} \sim
10^{15}$ numerical nodes is needed for every time step. The total
number of time steps must be also very large to be sure that the
magnetic vortices do not destroy the core of the magnetic obstacle.
In the previous Section, $Re=900$ and $N=9$, and the core of the
obstacle is shown to be stable.

Another interesting issue is whether the vortex shedding past the
magnetic obstacle is similar to that past the solid obstacle.
Indeed, the following is valid generally in both cases: (1) the
attached vortices are formed from the creeping flow when $Re$
prevails a critical value; (2) in a certain range of $Re$, the
attached vortices are stable; (3) when the inertial force exceeds
the stable threshold, the attached vortices detach from the body.
Because the inertial force is decisive in both cases and the
Lorentz force vanishes past the magnetic obstacle, it is expected
that the vortex shedding in both cases might be similar as well,
at least for specially selected geometrical conditions. This issue
is open currently.

\section*{Conclusion}

In this paper, we attempted to shed light on the peculiarities of
the MHD flow passing over a magnetic obstacle when the magnetic
interaction parameter $N$ is large, i.e. strength of the magnetic
field is high or when the Reynolds number $Re$, i.e. inertia of
the flow, is large. The corresponding case for moderate $Re$ and
$N$ has been elaborated in
\cite{Votyakov:PRL:2007,Votyakov:JFM:2007}. As it turns out high
values of $Re$ and $N$ neatly emphasize analogies between a
magnetic and a solid obstacle \bchange{that have been under
discussion from the beginning of MHD in the former USSR}. In this
paper, we have illustrated, by means of 3D numerical simulations,
how the core of the magnetic obstacle is formed when $N$ increases
and examined the shedding of attached vortices when the  $Re$ is
high enough.

With regard to the core of the magnetic obstacle the open issue
remains whether is possible to treat the problem in a simpler way
based on the Kulikovskii's approach \cite{Kulikovskii:1968}. This
is an old and fruitful idea to subdivide the complicated MHD flow
into two parts: core and periphery and then combine both parts with
appropriate boundary conditions. It is shown in this paper that
there must be three parts to include recirculation at high $N$: the
rest of the flow, immovable core, and transitional region with
magnetic vortices.

With regard to the shedding of vortices past a magnetic obstacle,
it is confirmed  that magnetic and connected vortices altogether
represent a virtual bluff body, which is spatially fixed owing to
the Hartmann friction \cite{Votyakov:PRL:2007,Votyakov:JFM:2007}.
Because the virtual body is fixed, a stagnancy region is formed.
In this region, the intensity of the magnetic field is negligible,
hence, the Lorentz force vanishes and attached vortices are
controlled only by the inertial force. The same happens past a
solid obstacle. Then, it may appear that the regularities known
for the attached vortices past a solid cylinder are valid also for
those past a magnetic obstacle. This way of thinking is useful
provided that one takes into consideration the
three-dimensionality of the problem, namely, the fact that the
cylinder is confined between two endplates.

\section*{Acknowledgements}

This work has been performed under the UCY-CompSci project, a
Marie Curie Transfer of Knowledge (TOK-DEV) grant (contract No.
MTKD-CT-2004-014199) funded by the CEC under the 6th Framework
Program. The simulations were carried out partially on a JUMP
supercomputer, access to which was provided by the John von
Neumann Institute (NIC) at the Forschungszentrum J\"{u}lich.
E.V.V. is grateful for many fruitful discussions with Oleg
Andreev, Yuri Kolesnikov, Andre Thess, and Egbert Zienicke during
his time in the Ilmenau University of Technology. \bchange{The
authors are thankful to the Referee whose comments led them to write a
section for the generic scenario of the wake of the magnetic
obstacle.}

\bibliographystyle{unsrt}
\bibliography{./../../Bibtex/mhd}

\end{document}